# Field-controlled domain wall pinning-depinning effects in ferromagnetic nanowire-nanoparticles system

V.L.Mironov, O.L.Ermolaeva, E.V.Skorohodov, and A.Yu.Klimov

Institute for physics of microstructures RAS, 603950, Nizhniy Novgorod, GSP-105, Russia
e-mail: mironov@ipm.sci-nnov.ru

We present the results of micromagnetic modeling and experimental investigations of field-driven domain wall (DW) pinning-depinning effects in the planar system consisting of a ferromagnetic nanowire (NW) and two ferromagnetic single domain nanoparticles (NPs). It was demonstrated that the magnitude of depinning field strongly depends on the spatial configuration of magnetic moments in the NPs subsystem. The algorithm of external magnetic field commutation and independent switching of NPs moments that permits the realization of magnetic logical calculations is discussed.

**Introduction**

The field-driven motion and pinning of domain walls in ferromagnetic nanowires are the subjects of intensive research motivated by promising applications for the development of magnetic logic and data storage systems [1-4]. The information in these devices is encoded as a direction of local magnetic moment in NW and write/erase or calculating processes are connected with domains reorientation accompanied by DWs nucleation, motion and annihilation. One of the main parameters limiting the operation rate of such systems is the velocity of DW propagation. This parameter is strongly dependent on the transverse field magnitude, DW structure, NW shape and dimensions [5-10]. In practice DW velocity reaches a value up to 1 km/s [10, 11]. On the other hand, the operation of DW based magnetic logical cells and data storage systems requires the controlled DW pinning for preservation from accidental data erasing and to save the results of intermediate calculations. The simple method of DW pinning is using the geometrical features in NW topology and artificially patterned traps at the NW edge [12-17]. The very interesting variants of DW pinning based on magnetostatic interaction of DW with superparamagnetic nanoparticle and with NW stubs were considered recently in Refs 18-20. In particular, it was shown that the system of several NW stubs enables the spatial configuring of local magnetic field and the controlling of DW pinning [20].

In this paper we consider an alternative combined system consisting of a ferromagnetic NW and two elongated single domain ferromagnetic NPs placed perpendicularly on either side of the NW. We report the results of micromagnetic modeling and experimental investigations of magnetization reversal and DW pinning-depinning effects in polycrystalline $Co_{60}Fe_{40}$ planar NW-NPs system and the dependence on the spatial configuration of magnetization in the NPs subsystem.

**Theoretical consideration**

We investigated the magnetization reversal processes in the NW-NPs system, which consists of nanowire with a special circular part (*N*) at one end and two NPs placed on either side of NW (see Fig. 1). The low-coercive circular pad N is used for the nucleation of domains with opposite orientation in an external magnetic field [21-24], while the NPs are used as a magnetic gate for the field-controlled DW pinning-depinning.

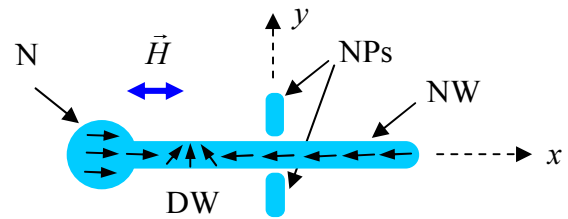

**FIG. 1.** The schematic drawing of the field-driven NW-NPs system.

To study the features of the pinning-depinning effects in the NW-NPs system we performed the computer micromagnetic simulations using the standard OOMMF (Object Oriented MicroMagnetic Framework) code [25]. We considered a system with the following parameters: NW and NP thickness was 20 nm; the NW width was 100 nm and NW length was 3 μm; the pad diameter was 200 nm; the lateral sizes of rectangular NPs were 100 × 200 nm; the NP-NW separation was 100 nm. The geometrical sizes of NW-NPs system were chosen close to the experimental structure described below. The calculations were carried out for the following CoFe parameters: the constant of exchange interaction was $J = 3 \cdot 10^{-6}$ erg/cm, the saturation magnetization was $M_S = 1900$ emu/cm$^3$ and the damping constant was 0.5 [26]. We omitted magnetocrystalline anisotropy assuming the polycrystalline structure of our samples. In model calculations the NW-NPs system was discretized into rectangular parallelepipeds with a square base of size $\delta = 5$ nm in the $x, y$ plane and height $h = 20$ nm.

The preliminary simulations showed that the coercive field (in $x$ direction) for NW without pad $H_{NW}$ was 500 Oe; the DW nucleating field for NW with pad $H_{Nuc}$ was 250 Oe; the NPs coercive field in $x$ direction was $H_{NP}^x = 1200$ Oe; the NPs coercive field in y direction for was $H_{NP}^y = 800$ Oe (the steps of simulated external field were 10 Oe). Note that the NP's coercive field along the $x$ direction is much larger than the NW's coercive field $H_{NP}^x > H_{NW}$.

We performed the micromagnetic simulations of pinning-depinning processes in the NW-NPs system. The computer experiment had the following scenario. At the first stage, the NW was magnetized uniformly (from right to left $-x$ direction) in an external magnetic field $H_{NP}^x > H_{ex} > H_{NW}$. Afterwards we applied the reversed field (from left to right $+x$ direction) $H_{NW} > H_{ex} > H_{Nuc}$. In this case the reorientation of magnetization in the circular pad and transverse DW formation were observed. The result of micromagnetic simulation of the DW in the NW is represented in Fig. 2. It is seen that the DW had the characteristic trapezoidal structure.

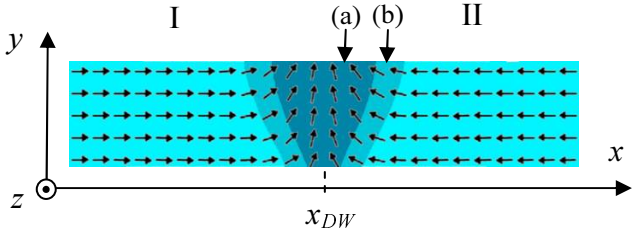

**FIG. 2.** The simulated DW structure in $100 \times 3000 \times 20$ nm CoFe NW. The quasi-uniform regions are indicated as I and II. The magnetization vectors in the region (a) are turned on the angles $\alpha_a \geq 0.5\,\pi/2$ and in the region (b) on $0.1\,\pi/2 \leq \alpha_b \leq 0.5\,\pi/2$. The regions (a) and (b) are marked by dark tones.

The process of NW magnetization reversal is accompanied by DW propagation from the pad to the free NW end and can be stopped due to magnetostatic interaction with the NPs gate. To estimate the pinning energy and depinning fields for different configurations of magnetization in the NW-NPs system we considered the dependence of the DW-NPs interaction energy on DW position. In general, the potential energy of the NW in an external magnetic field can be calculated as follows:

$$E_{NW}(x_{DW}) = -\int_{V_{NW}} \vec{M}_{NW} \vec{H}\, dV, \quad (1)$$

where $\vec{M}_{NW}(\vec{r})$ is the magnetization distribution in the NW, $\vec{H}$ is the sum of magnetic stray fields from the NPs and the uniform external magnetic field $\vec{H} = \vec{H}_{NP} + \vec{H}_{ex}$; $x_{DW}$ is the position of the DW center; the integration is performed over the NW volume. In estimating calculations, we supposed that the NPs are magnetized uniformly and the magnetization of the NW and NPs does not depend on the $z$ coordinate. Then $x$ and $y$ components of a nanoparticle's stray magnetic field can be written in the following form:

$$H_{NP\,x}(x,y) = M_S \int_{V_{NP}} \frac{3(y+y_{NP})(x+x_{NP})}{\left((y+y_{NP})^2 + (x+x_{NP})^2\right)^{5/2}} dV_{NP}, \quad (2)$$

$$H_{NP\,y}(x,y) = M_S \int_{V_{NP}} \frac{2(y+y_{NP})^2 - (x+x_{NP})^2}{\left((y+y_{NP})^2 + (x+x_{NP})^2\right)^{5/2}} dV_{NP}, \quad (3)$$

where the integration is performed over the nanoparticle volume $V_{NP}$, $x_{NP}$ and $y_{NP}$ are the coordinates of integration.

The dependence of the interaction energy $E_{NW}$ on DW position allows one to estimate the pinning energy and depinning fields for different configurations of magnetization in the NW-NPs system. We carried out the numerical calculations of energy landscape based on formulas (1)-(3) and the model of rigid DW, omitting the changing of exchange and demagnetizing energy with DW propagation from pad to the free NW end [20, 27]. We used the simulated DW magnetization distribution (Fig. 2) as the model distribution in the $E_{NW}(x_{DW})$ calculations.

The effects of DW pinning depend on the mutual configuration of magnetization in the NW and NPs. Different possible variants of NW and NPs magnetization, corresponding dependences of NW-NPs system energy on DW position and results of micromagnetic modeling are represented in Figures 3 - 6.

Let us analyze firstly the configuration represented in Fig. 3a (A-type configuration). In this case the magnetization vectors in NW and in NPs are directed towards each other (head-to-head or tail-to-tail configurations). The corresponding dependences of energy $E_{NW}$ on DW position at different external fields are represented in Fig. 3b. It is seen that DW propagation is connected with overcoming the energy barrier, which is defined mainly by magnetostatic interaction of NW parts I and II with NPs field ($x$ component). Note that the magnitude of barrier is independent from the DW direction due to symmetry of NPs magnetic configuration. Hence, in a weak magnetic field DW will be pinned in the region before NPs gate. The estimate of energy barrier at zero field is $E_B = 8.2 \cdot 10^{-10}$ erg. In an external magnetic field the pinning barrier is decreased and at 470 Oe (depinning field $H_B$) it vanishes completely (curve 2 in Fig. 3b). As it was obtained in our comparative accurate micromagnetic simulations taking into account the effects of DW and NPs magnetization disturbance, the relative error in $E\ (x\ )$ calculations based on rigid DW model is less than 5 % (see dots at the curve 1 in Fig. 3b) and can be neglected in qualitative considerations. The micromagnetic simulations also confirmed the DW pinning before the NPs gate. The position of pinned DW in the external magnetic field $H_{ex} = H_{Nuc}$ is shown in Fig. 3c. The critical depinning field estimated directly from OOMMF modeling were $H\ = 480$ Oe (in simulation the field increasing step was 10 Oe ).

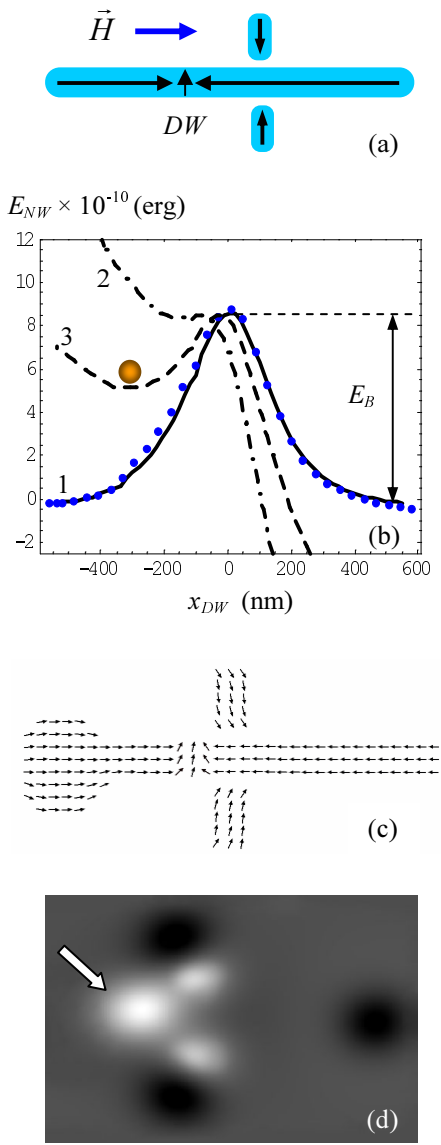
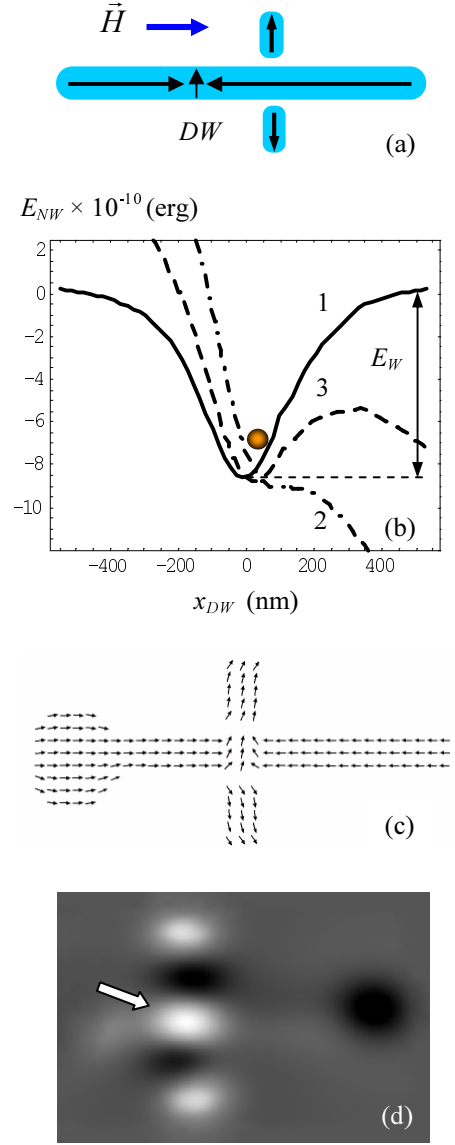

**FIG. 3.** (a) is the A-type configuration of magnetization in the NW-NPs system in an external magnetic field. (b) is the potential energy profiles $E_{NW}(x_{DW})$ for different external magnetic fields. The solid line (1) is the energy profile at zero field. The dot-dashed line (2) is for the critical external field $H_B = 470$ Oe. The dashed line (3) is for the intermediate field $0.5\,H_B$. The DW pinning position is indicated schematically by circle at curve (3). (c) is the model magnetization distribution at $H = H_{Nuc} = 250$ Oe demonstrating the DW pinning on the potential barrier before the NPs. (d) is the model MFM contrast distribution from the NW-NP system (without nucleating part) corresponding to the magnetization distribution shown in Fig. 3c. The MFM pole from the transverse DW is indicated by the white arrow.

**FIG. 4.** (a) is the B-type configuration of magnetization in NW-NPs system in an external magnetic field. (b) is the energy $E_{NW}$ profile for different external magnetic fields. The solid line (1) is the energy profile at zero external field. The dot-dashed line (2) is for the critical magnetic field $H_W = 470$ Oe. The dashed line (3) is for the intermediate field $0.5\,H_W$. The DW pinning position is indicated schematically by circle. (c) is the model magnetization distribution at $H = H_{Nuc} = 250$ Oe demonstrating the DW pinning in the potential well between NPs. (d) is the MFM contrast distribution from the NW-NP system (without nucleating part) corresponding to the magnetization distribution shown in Fig. 4c. The MFM pole from transverse DW is indicated by white arrow.

Besides, we calculated the expected MFM contrast distributions for the configuration of magnetization presented in Fig. 3c to compare with results of experimental MFM investigations. The model MFM image was calculated as the phase shift $\Delta\varphi$, of cantilever oscillations under the gradient of the magnetic force [28, 29]:

$$\Delta\varphi = -\frac{Q}{K}\frac{\partial F_z}{\partial z} \sim \frac{\partial^2 H_z}{\partial z^2}, \qquad (4)$$

where $Q$ is the cantilever quality factor, $K$ is the cantilever force constant, and $F_z$ is the z-component of magnetic force. This value is proportional to the second derivative of sample magnetic stray field $\partial^2 H_z/\partial z^2$. The model MFM image corresponding to the model magnetization distribution (Fig. 3c) is presented in Fig. 3d. The calculations were performed for the scanning height of 50 nm (close to experimental conditions). The DW is seen as the bright pole located before the NPs poles (it is indicated by arrow in Fig. 3d).

In the B-type configuration represented in Fig. 4a we have head-to-head (tail-to-tail) magnetization in the NW

but tail-to-tail (head-to-head) magnetization in the NPs. The energy profile $E_{NW}(x_{DW})$ has the potential well (Fig. 4b) and the DW is pinned at the region directly between the NPs. Thus, the NW remagnetization is connected with DW escape from the energy well, which is defined mainly by magnetostatic interaction of NW parts I and II with NPs field ($x$ component) and does not depend on DW direction. The estimate of the activation energy $E_w$ at the zero field is $8.2 \cdot 10^{-10}$ erg. In an external magnetic field the pinning barrier connected with the potential well is decreased (see curve 3 in Fig. 4b) and at 470 Oe (depinning field $H_W$) it vanishes completely (curve 2 in Fig. 4b).

The micromagnetic modeling confirmed the DW pinning between NPs for the B-type configuration. The magnetization distribution in the NW-NPs system with pinned DW is represented in Fig. 4c. The depinning field estimated directly from the OOMMF simulations was 480 Oe.

The Fig. 4d shows the expected MFM contrast distribution corresponding to the configuration of magnetization presented in Fig. 4c. The DW is seen as the bright pole located between the NPs poles (it is indicated by arrow in Fig. 4d).

The third possible configuration, in which the NPs moments have the same direction (head-to-tail) but the DW magnetization has the opposite direction (C-type configuration), is presented in Fig. 5a. In this case the $x$ components of stray magnetic fields from the NPs are partly compensated. The calculated energy landscape has the potential barrier caused by interaction of the DW with the $y$ component of the NPs field (Fig. 5b). Hence, in the C-type configuration a DW can be pinned in the region before NPs. The model distribution of magnetization for the C-type configuration with pinned DW is presented in Fig. 5c. In simulations we artificially initialized a DW in the left front of the NW and used the small driving field $H$ = 25 Oe to stimulate DW propagation. The effect of DW pinning was registered as the stabilization of DW position in the external magnetic field. The C-type configuration has a very small pinning energy in comparison with A configurations. The estimate of the energy barrier $E_b$ at zero external field is $1.05 \cdot 10^{-10}$ erg. The calculated depinning field (estimated as barrier vanishing in the external field) is 90 Oe that practically coincides with 100 Oe estimated directly from the OOMMF simulations.

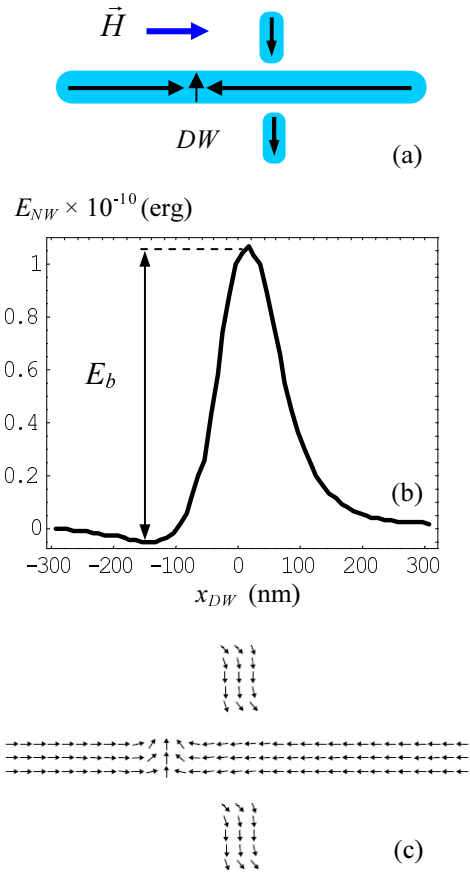
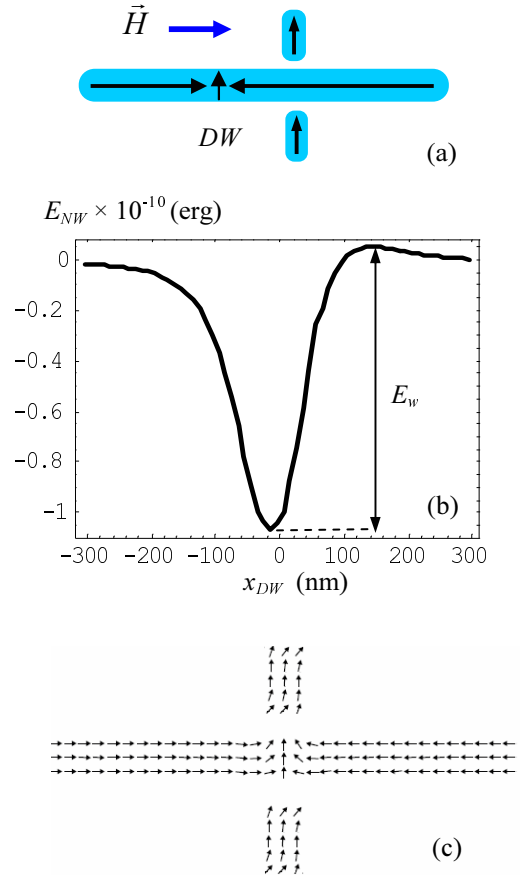

**FIG. 5.** (a) is the C-type configuration of magnetization in NW-NPs system. (b) is the energy profile $E_{NW}(x_{DW})$ at zero external field. (c) is the model magnetization distribution at zero external field (after applying driving field $H$ = 25 Oe) demonstrating the DW pinning on the potential barrier before the NPs.

**FIG. 6.** (a) is the D-type configuration of magnetization in NW-NPs system. (b) is the energy profile $E_{NW}(x_{DW})$ at zero external field. (c) is the model magnetization distribution at zero external field (after applying $H$ = 25 Oe) demonstrating the DW pinning in the potential well between NPs.

The fourth possible configuration (D-type) is presented in Fig. 6a. In this case the directions of magnetization in both NPs are the same (head-to-tail) and coincide with DW orientation. The energy landscape $E_{NW}(x_{DW})$ has the potential well caused by interaction of the DW with the *y* component of the NPs field (Fig. 6b). Hence, in this case the DW can be pinned in the region directly between the NPs (see the position of pinned DW in Fig. 6c) but this configuration has a small pinning energy in comparison with B configurations. The estimate of the energy barrier $E_w$ at zero field is $1.05 \cdot 10^{-10}$ erg. The depinning field estimated as the potential wall vanishing in the external field is 90 Oe. The depinning field estimated directly from the OOMMF simulations was 100 Oe.

### Experimental

The NW-NPs systems were fabricated using a negative e-beam lithography and ion etching processes. The $Co_{60}Fe_{40}$ (20 nm) / V (15 nm) / Cu (10 nm) multilayer structure was deposited onto a Si substrate by magnetron sputtering. Afterward the sample was covered by fullerene $C_{60}$ (80 nm), which was used as e-beam resist. The initial protective mask was formed in $C_{60}$ by exposure in the ELPHY PLUS system (based on the scanning electron microscope "SUPRA 50VP") with subsequent chemical treatment in an organic solvent. Afterward, the image was transferred in Cu layer by $Ar^+$-ion etching and further in V layer by plasma etching in Freon. At the final stage, the NW-NPs system was fabricated in ferromagnetic $Co_{60}Fe_{40}$ layer by $Ar^+$-ion etching. The characteristic scanning electron microscope (SEM) image of the NW-NPs system is represented in Fig. 7.

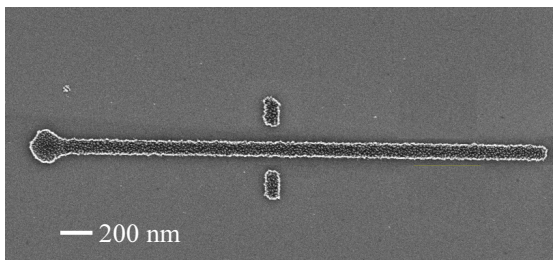

FIG. 7. The SEM image of nanowire-nanoparticles system.

The width of the NW was 100 nm, the NW length was about 2.8 μm; lateral dimensions of the NPs were $100 \times 200$ nm, the NP-NW separation was 100 nm; the diameter of nucleating part was 200 nm.

The magnetic states and the magnetization-reversal effects in the NW-NPs system were studied using a vacuum multimode scanning probe microscope "Solver-HV," which is equipped with a dc electromagnet incorporated in a vacuum vibration insulating platform (the maximal magnitude of magnetic field is 1 kOe). The scanning probes were cobalt coated with a thickness of 30 nm. Before measurements, the tips were magnetized along the symmetry axes (*Z*) in a 10 kOe external magnetic field.

The magnetic force microscope (MFM) measurements were performed in the non-contact constant height mode. The phase shift $\Delta \varphi$, of cantilever oscillations under the gradient of the magnetic force was registered to obtain the MFM contrast. All measurements were performed in a vacuum of $10^{-4}$ Torr, which improved the MFM signal due to an increase in the cantilever quality factor.

### Results and discussion

We investigated the dependence of the pinning-depinning processes in the NW-NPs system (Fig. 7) on the configuration of magnetization in the NPs. The experiment was performed in situ in the MFM "Solver HV" vacuum chamber. First, we studied the NW magnetization reversal when magnetic moments in NW-NPs system corresponded to the A-type configuration. The different stages of NW remagnetization experiment are represented in Fig. 8. The initial state was prepared by previous sample magnetizing along the NW (in − *x* direction) in $H_{||} \geq 700$ Oe (approximate rate of field increasing was 100 Oe/s) and subsequent NPs magnetizing in perpendicular (in *y* direction) magnetic field $H_\perp = 700$ Oe and remagnetizing (in − *y* direction) in reversed $H_\perp$ with magnitude of about 490 Oe. The head to head configuration is formed in reversed field due to small difference in NPs coercivity. The MFM image of the initial state is presented in Fig. 8a. It is seen that the MFM contrast poles positions at the MFM image of the initial state correspond to the type A configuration (the dark pole corresponds to the tail and bright pole corresponds to the head of the magnetization vector). Afterwards the external field $H_{||}$ was applied in the reversed (*x*) direction and NW remagnetization effect was studied. All external field manipulations were performed with 10 Oe steps. An approximate rate of change of the field was about 100 Oe/s. When $H_{||}$ exceeded 300 Oe (DW nucleation field $H_{Nuc}$) we registered the appearance of an additional bright pole in the MFM image (indicated by the white arrow in Fig. 8b) corresponding to the DW (compare with Fig. 3d). So we observed the DW pinning on the potential barrier before the NPs. The DW position was stable in external magnetic fields up to $H_{||} = 560$ Oe, but when $H_{||}$ exceeded 560 Oe (DW depinning field $H_B$) the remagnetization of the NW was observed (see changing of MFM contrast at the free NW end in Fig. 8c).

The different situation was observed for the magnetization reversal experiment in B-type configuration of NW-NPs magnetic moments. The initial state was prepared by an analogous method to that of the A-type configuration but for generation of tail-to-tail configuration in NPs we used the inversed procedure of NPs remagnetization in a perpendicular magnetic field. The different stages of the NW remagnetization experiment are demonstrated in Fig. 9. The MFM image of the initial state is presented in Fig. 9a. The MFM poles positions confirm the B-type configuration. Analogously when the external magnetic field $H_{||}$ (*x* direction) exceeded 300 Oe we registered the appearance of the additional bright MFM pole (indicated by white arrow in Fig. 9b) corresponding to the DW. But it is seen that in this case the DW was pinned right in between NPs.

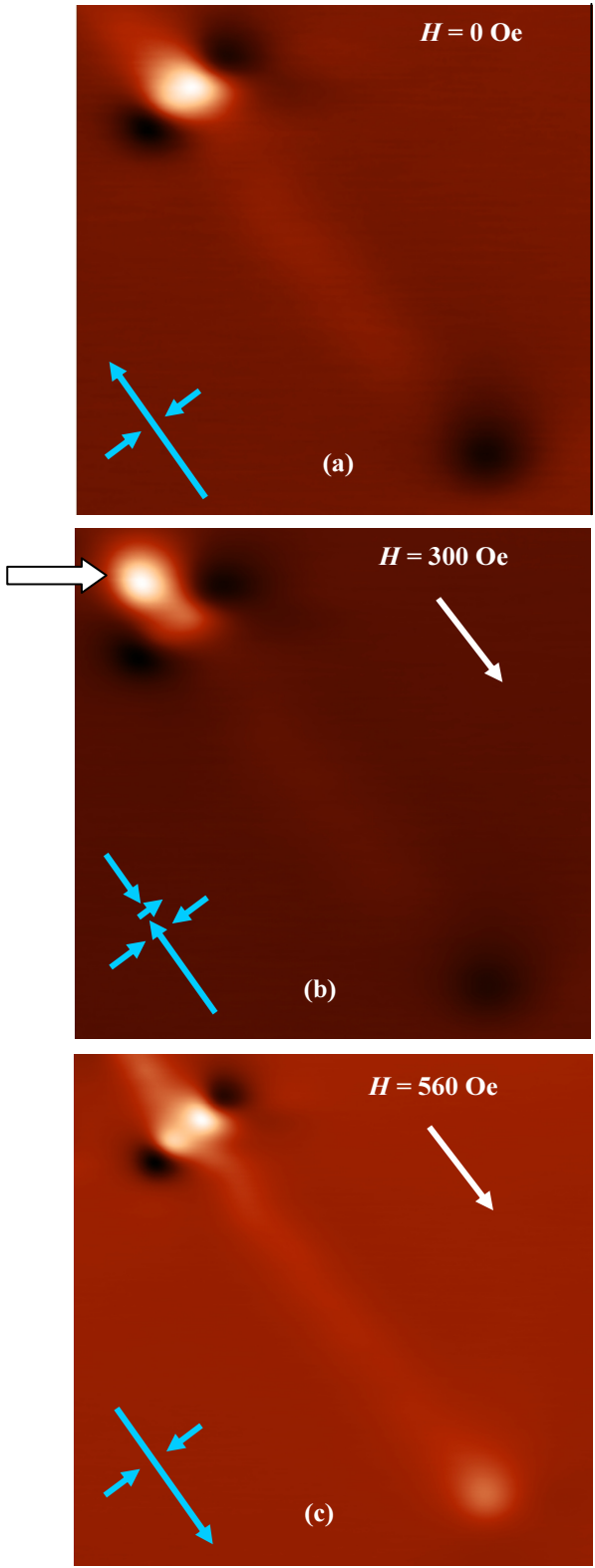
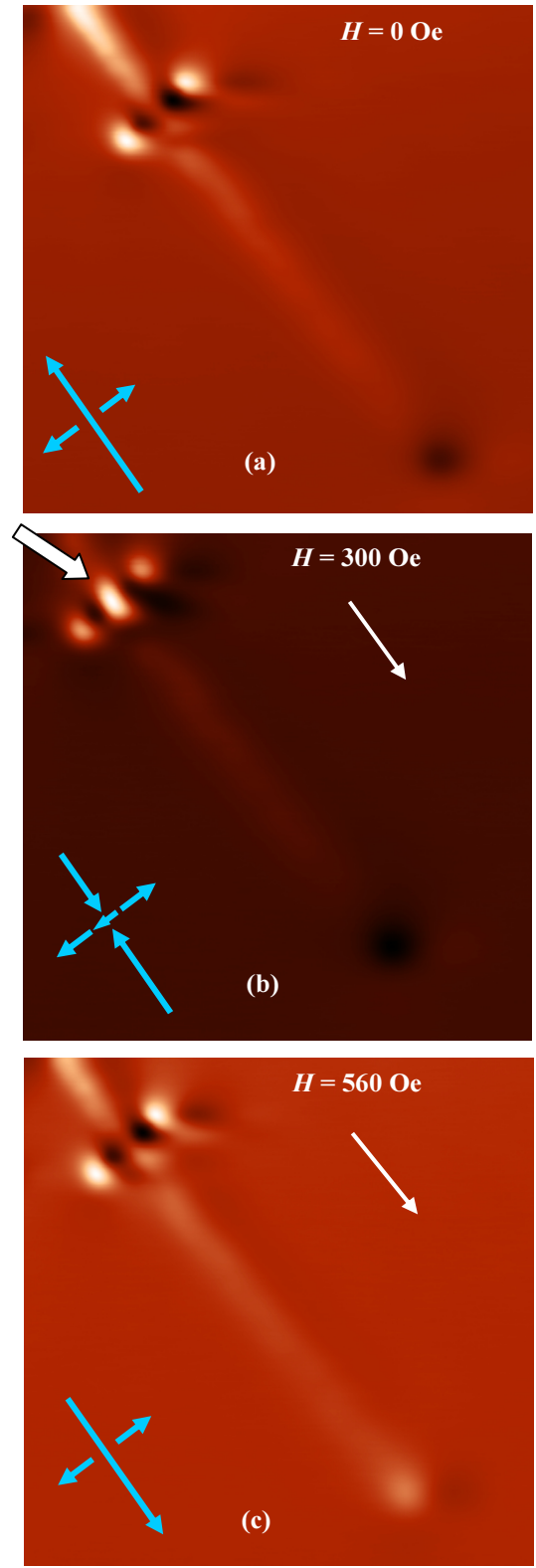

**FIG. 8.** The MFM images of NW-NPs system after the application of an external magnetic field. (a) is initial state with A-type configuration after previous magnetizing. (b) is MFM image of DW pinned at the barrier near the NPs after applying 300 Oe external field (DW position is indicated by white arrow). (c) is MFM image of the NW-NPs system after remagnetization in 560 Oe external magnetic field. The schemes of magnetization are represented at left bottom corners.

**FIG. 9.** The MFM images of NW-NPs system after the application of an external magnetic field. (a) is initial state with B-type configuration after previous magnetizing. (b) is MFM image of DW pinned at the well between NPs after applying 300 Oe external field (DW position is indicated by white arrow). (c) is MFM image the NW-NPs system after remagnetization in 560 Oe external magnetic field. The schemes of magnetization are represented at left bottom corners.

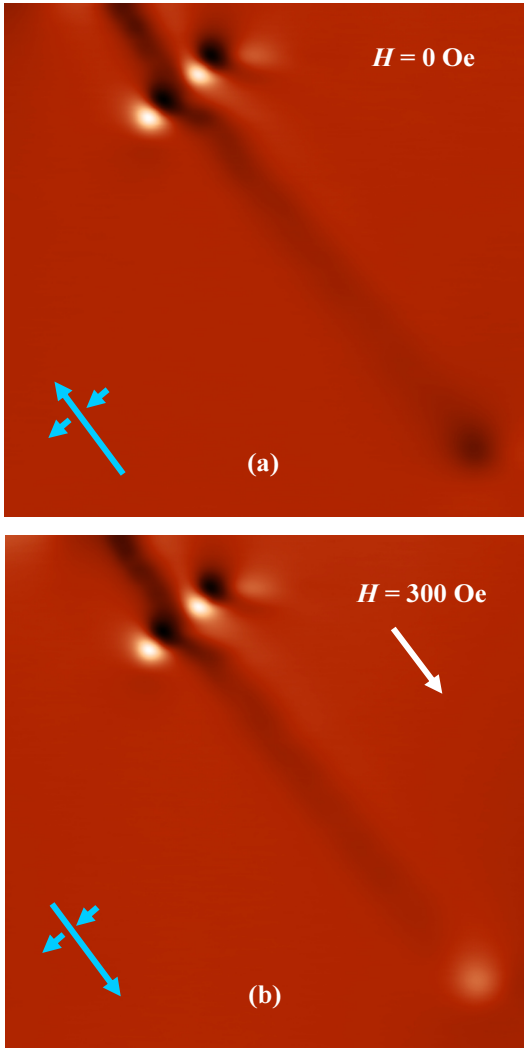

FIG. 10. The MFM images of magnetization reversal of NW-NPs system with C(D)-type configuration in an external magnetic field. (a) is initial state after previous magnetizing. (b) is MFM image of NW-NPs system after applying 300 Oe external field. The DW pinning was not observed because for this system the pinning field was much less than nucleating field.

We believe that it can be explained as DW pinning in a potential well (compare with Fig. 4d). The depinning field was practically the same as in the case of A-type configuration, so when $H_{||}$ exceeded 560 Oe (DW depinning field $H$ ) we observed the remagnetization of the NW, which was registered as a changing of MFM contrast on the free NW end (see Fig. 9c). The similarity between the magnitudes of depinning fields $H_B$ and $H$ in A- and B- type configurations observed in model simulations as in experimental measurements allows one to suggest that the magnetostatic DW – NPs interaction in such system is quite weak and the rigid DW approximation is valid in this case

The results of NW magnetization reversal in NW-NPs system with C (or D)-type configuration are presented in Fig. 10. The initial state was prepared by previous magnetizing along the NW (– x) and subsequent NPs remagnetization in perpendicular magnetic field (– y) with magnitude of 700 Oe exceeding the NP's coercivity.

Subsequently the reversed external magnetic field $H_{||}$ was applied and the process of remagnetization was studied. In this case we did not registered the pinning of the DW since the remagnetization effect was observed just after applying the nucleating external magnetic field $H_{||} = H$ = 300 Oe. This fact demonstrates that for our NW-NPs system the DW nucleation field is larger than the depinning field connected with DW-NPs interaction in C(D)-type configuration ( $H_{Nuc} > H_b, H_w$ ).

As it is seen, the experimental results are in qualitative accordance with the previous theoretical estimating and micromagnetic modeling. However, there are considerable quantitative differences in estimations of $H_{Nuc}$, $H$ and $H$ fields. We believe that the high values of nucleating and depinning fields observed in the experiments are connected with NW edge roughness. As it was estimated from SEM images the RMS edge roughness for our NW-NPs system was about 8 nm (8 % of width). The direct OOMMF simulations taking into account the estimated edge roughness have demonstrated the considerable increasing of nucleating and depinning fields close to the experimentally observed values.

The described above NW-NPs system can be considered as a prototype of a magnetic logical cell (MLC), which realizes the "exclusive disjunction" logical operation (so-called XOR). For example, the input signals can be encoded as a direction of magnetic moment in NPs and output information as a moment direction at free end of NW. The Fig. 11 shows a scheme of inputs-output and possible input-output information coding.

Table 1.

The logical output states for all input states (truth table).

| Input 1 | Input 2 | Output |
|---------|---------|--------|
| 0 | 0 | 0 |
| 0 | 1 | 1 |
| 1 | 0 | 1 |
| 1 | 1 | 0 |

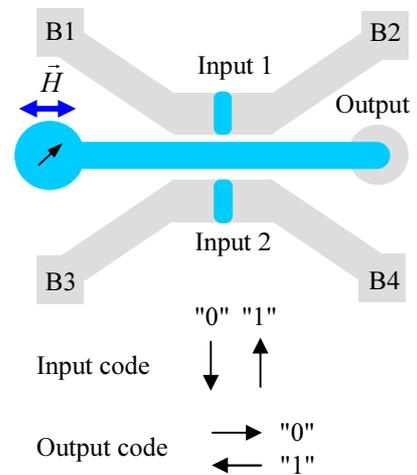

FIG. 11. The schematic NW-NPs logical cell drawing and the input-output states coding.

The algorithm of MLC operation includes the periodic reactivation cycle and the logical computing cycle consisting of input information writing and reading of the result of logical operation. The logical calculations can be organized as follows. A first stage is the initialization process when the logical "1" is written in the NW by an external magnetic field $H_\parallel$ ($H_\parallel > H_B, H_W$) applied in $-x$ direction. Afterwards the input information is written in the NPs by local magnetic fields, which can be created for instance with current buses B1-B2 and B3-B4 (see Fig. 11). At the final stage the reversed external magnetic field $H_\parallel$ with amplitude $H_B, H_W > H_\parallel > H_{Nuc}$ ($H_{Nuc} > H_b, H_w$) is applied in the $x$ direction and the output information is read. The magnetic state of the NW free end can be analyzed using local magneto optical Kerr effect or by means of tunneling magnetoresistance element. Afterwards the cycle of operation is repeated. The correspondence between input and output information in such a MLC is represented in the Table 1.

**Conclusion**

Thus, we investigated the DW pinning-depinning effects in a hybrid system consisting of a ferromagnetic NW (with circular pad at one end) and a two-NP gate placed perpendicularly at the middle part of NW.

The micromagnetic simulations and direct magnetic for ce microscopy measurements have showed that in dependence on relative orientation of magnetic moments in NW and NPs subsystem there are two variants of DW pinning connected with a potential barrier (A-type configuration) or a potential well (B-type configuration)caused by magnetostatic interaction between the DW magnetization structure in the NW and local NPs stray field. For the $Co_{60}Fe_{40}$ based NW-NPs system consisting of 100 × 2800 × 20 nm nanowire with 200 nm in diameter nucleating part and 200 × 100 × 20 nm nanoparticles gate (with 100 nm NW-NP separation) the nucleating field 300 Oe and depinning field 560 Oe in A-type and B-type configurations were registered. When the magnetic moments of the NPs were set in C-type or D-type configurations, the DW pinning effects were not observed at all.

Potentially the different combinations of NWs with independently switched NPs are very promising for the development of new type magnetic logic cells and other DW based computing systems.

**Acknowledgements**

The authors are very thankful to S. A. Gusev, S. N. Vdovichev and V. V. Rogov for assistance in samples preparation and to B. A. Gribkov and A. A. Fraerman for the very fruitful discussions. We are grateful to Prof. J. A. Blackman (University of Reading, U.K.) for useful discussions and improvement of English.

This work was supported by the Russian Foundation for Basic Research (projects 11-02-00434 and 11-02-00589). E.V.S wishes to thank Carl Zeiss company for financial support.